\documentclass[10pt,english,aps,prd,floatfix,showpacs,nofootinbib,notitlepage]{revtex4-1}
\usepackage{graphicx,xcolor}
\usepackage[colorlinks=true,linktocpage=true,linkcolor=blue,citecolor=blue,allcolors=blue]{hyperref}
\usepackage{amsmath}

\usepackage{soul} 
\usepackage[utf8]{inputenc}  

\newcommand{\be}{\begin{eqnarray}}
\newcommand{\ee}{\end{eqnarray}}
\newcommand{\non}{\nonumber\\}

\newcommand{\ave}[1]{\left\langle #1 \right\rangle}

\newcommand{\qp}{Q^{+}}
\newcommand{\qpm}{q^{+}}
\newcommand{\qm}{Q^{-}}
\newcommand{\qmm}{q^{-}}
\newcommand{\qh}{\hat{Q}^{+}}
\newcommand{\qhm}{\hat{q}^{+}}
\newcommand{\et}{\tilde{\epsilon}_p}
\newcommand{\eq}{\epsilon_q}
\newcommand{\eqb}{\epsilon_{\bar{q}}}
\newcommand{\ep}{\epsilon_p}
\newcommand{\epb}{\epsilon_{\bar{p}}}
\newcommand{\Np}{N_p}
\newcommand{\Npb}{N_{\bar{p}}}
\newcommand{\npm}{n_p}
\newcommand{\nbm}{n_{\bar{p}}}
\newcommand{\npt}{\tilde{n}_p}
\newcommand{\zp}{z_p}
\newcommand{\zq}{z_q}
\newcommand{\Zq}{Z_Q}
\newcommand{\Zp}{Z_P}
\newcommand{\Cov}[1]{\Sigma^{1,1}_{#1} }
\newcommand{\cov}[1]{\sigma^{1,1}_{#1}}
\newcommand{\FF}[2]{F^{#1}_{#2}}
\newcommand{\ff}[2]{f^{#1}_{#2}}

\newcommand{\eqn}[1]{\begin{align} #1 \end{align}}

\begin{document}
\setstcolor{red} 

\title{Efficiency corrections for factorial moments and cumulants of overlapping sets of particles
}

\author{
Volodymyr Vovchenko
}
\email[E-Mail:]{vovchenko@lbl.gov}
\affiliation{
Nuclear Science Division,
Lawrence Berkeley National Laboratory, 1 Cyclotron Road,
Berkeley, CA 94720
}
\author{
  Volker Koch
}
\email[E-Mail:]{vkoch@lbl.gov}
\affiliation{
Nuclear Science Division,
Lawrence Berkeley National Laboratory, 1 Cyclotron Road,
Berkeley, CA 94720
}

\date{\today}

\begin{abstract}
  In this note we discuss subtleties associated with the efficiency corrections for measurements of off-diagonal cumulants and factorial moments for a situation when one deals with overlapping sets of particles, such as correlations between numbers of protons and positively charged particles.
  In particular, we discuss the situation commonly encountered in heavy-ion experiments, where first all charges are reconstructed and then protons are selected from these charges by an additional particle identification procedure.
  We present the efficiency correction formulas for the case when the
  detection efficiencies follow a binomial distribution. 
\end{abstract}

\maketitle

\section{Introduction}

Fluctuations of conserved charges as a probe of the phase structure of strongly interacting matter
have recently received considerable interest theoretically as well as experimentally (for  a recent
review, see \cite{Bzdak:2019pkr}). 
For example, higher order cumulants of the net baryon density are
sensitive to the existence of a critical point \cite{Stephanov:2008qz},
and may also provide insights about the chiral criticality governing the cross-over transition at vanishing baryochemical potential \cite{Morita:2013tu}. 
The so-called off-diagonal cumulants, i.e. correlations between
two different conserved charges, such as baryon number and strangeness, 
on the other hand, 
provide insight about the effective degrees of freedom in the medium~\cite{Koch:2005vg,Bazavov:2013dta}.

Experimentally, these cumulants are measured by analyzing event-by-event distributions of particles produced in heavy-ion collisions. 
For these measurements to reveal the true fluctuations of the system created in these collisions, one needs to take into account and remove fluctuations induced by the detector measurement process itself.
These detector induced fluctuations,
often referred to as efficiency fluctuations \cite{Kitazawa:2012at,Bzdak:2012ab,Bzdak:2013pha,Luo:2014rea}, 
arise from the finite detection probability $W_{D}(n,N)$ of an actual detector, where $W_{D}(n,N)$ is the probability to observe $n$ particles given $N\geq n$ particles in an event. 
The probability distribution of observed particles, 
$p(n)$, is related to the distribution of true particles, $P(N)$, by
\begin{align}
p(n) = \sum_{N}W_{D}(n,N) P(N)~.
\label{eq:p_n_general}
\end{align}
Consequently, the cumulants of the observed distribution $p(n)$ differ from those of the true
distribution, $P(N)$. 
Therefore, an unfolding procedure is needed, which mathematically corresponds to finding the inverse of $W_{D}(n,N)$. 
This is not an easy task in general~(see discussion e.g. in \cite{Bzdak:2016qdc}). 
However, if  $W_{D}(n,N)$ can be approximated by
a binomial distribution -- a reasonably good approximation in a number of cases (see
\cite{Adamczewski-Musch:2020slf,Adam:2020unf}) -- the relevant formulas for the efficiency corrections of cumulants can and have been derived \cite{Bzdak:2012ab,Bzdak:2013pha,Kitazawa:2016awu,Nonaka:2018mgw}.

However, certain subtleties arise when efficiency corrections are performed for off-diagonal cumulants, such as the correlation of net-proton or net-kaon number with the net-charge number.
These subtleties have not yet been addressed in the literature. 
It is the purpose of this note to discuss and provide the necessary efficiency correction formulas. 
These may be useful for the ongoing and future heavy-ion experiments, in particular as an effort to measure such off-diagonal cumulants is underway~(see e.g.~\cite{Adam:2019xmk}).

\section{A reminder on efficiency corrections for nonoverlapping sets of particles}

In the following we shall denote all true quantities with upper case letters and all measured quantities, which are affected by detector efficiencies, with lower case symbols.
Following  \cite{Bzdak:2012ab} it is convenient to express the cumulants in terms of factorial
moments, as efficiency corrections for those are simpler. 
For example the co-variance or off-diagonal cumulant
between two {\em distinct} particle species, $A$ and $B$~($A \cap B = \emptyset$), defined as
\begin{align}
  \Cov{A,B} = \ave{N_{A}N_{B}} - \ave{N_{A}}\ave{N_{B}},
\label{eq:offdiag_distinct}
\end{align}
may be repressed in terms of the factorial moments
\begin{align}
  \FF{i,j}{A,B}=\ave{\frac{N_{A}!}{(N_{A}-i)!}\frac{N_{B}!}{(N_{B}-j)!}}
  \label{eq:fac_cum}
\end{align}
so that 
\begin{align}
  \Cov{A,B} = \FF{1,1}{A,B}-\FF{1,0}{A,B}\FF{0,1}{A,B}.
\end{align}
The second moments are given by
\begin{align}
\ave{N_{A}^{2}}=\FF{2,0}{A,B}+\FF{1,0}{A,B}
\end{align}
and also the higher moments can be expressed as combinations of the factorial moments.

Given the multiplicity distribution function $P(N_{A},N_{B})$ for particles of type $A$ and $B$, the corresponding factorial cumulants are conveniently obtained through the generating function
\begin{align}
G(Z_{A},Z_{B})=\sum_{N_{A},N_{B}}Z_{A}^{N_{A}}Z_{B}^{N_{B}}P(N_{A},N_{B})
\end{align}
via
\begin{align}
F^{i,j}_{A,B}=\frac{d^{i+j}}{d Z_{A}^{i} \, dZ_{B}^{j} }\left. G(Z_{A},Z_{B})\right|_{Z_{A}=1,Z_{B}=1}~.
  \label{eq:facmom_from_gen}
\end{align}
Consider now a case when the  probability to detect a particle is governed by a binomial distribution,
\begin{align}
W(n,N)=B(n,N;\epsilon)=\frac{N!}{n!(N-n)!}\epsilon^{n}(1-\epsilon)^{(N-n)}
  \label{eq:binomial}
\end{align}
with $\epsilon$ being the so-called efficiency. 
The distribution of measured particles is then given by
\begin{align}
p(n_{A},n_{B})=\sum_{N_{A},N_{B}}B(n_{A},N_{A};\epsilon_{A}) B(n_{B},N_{B};\epsilon_{B}) P(N_{A},N_{B})~.
\end{align}
The resulting factorial moment generation function which provides the {\em measured} factorial
moments is then
\begin{align}
g(z_{A},z_{B}) & =\sum_{n_{A},n_{B}}z_{A}^{n_{A}}z_{B}^{n_{B}}p(n_{A},n_{B}) \nonumber \\
& =\sum_{N_{A},N_{B}}(1-\epsilon_{A}+z_{A}\epsilon_{A})^{N_{A}} (1-\epsilon_{B}+z_{B}\epsilon_{B})^{N_{B}}P(N_{A},N_{B}),
\end{align}
where we used the fact that  $\sum_{n=0}^{N}z^{n}B(n,N,\epsilon) = (1 - \epsilon + z \epsilon)^{N}$.
The measured factorial moments reduce simply to
\begin{align}
\ff{i,j}{A,B}=\epsilon_{A}^{i}\epsilon_{B}^{j}\FF{i,j}{A,B},
 \label{eq:f_ij_standard_1}
\end{align}
thus, the true factorial moments, $F_{i,j}^{A,B}$, can be recovered by dividing the measured ones by
the appropriate powers of the efficiencies
\begin{align}
\FF{i,j}{A,B} = \frac{\ff{i,j}{A,B}}{\epsilon_{A}^{i}\epsilon_{B}^{j}}, \qquad A \cap B = \emptyset.
  \label{eq:f_ij_standard}
\end{align}
As result the true co-variance or off-diagonal cumulant is given by
\begin{align}
\Cov{A,B}=\frac{\cov{A,B}}{\epsilon_{A}\epsilon_{B}}, \qquad A \cap B = \emptyset.
\label{eq:cov_ij_standard}
\end{align}
Therefore, as long as we have {\em different} particle species or even distinct groups of particles, such as
protons and pions, correcting cumulants for efficiency simply entails expressing the cumulants in
terms of factorial moments and then make use of Eq.~\eqref{eq:f_ij_standard}, as discussed in
detail in \cite{Bzdak:2012ab}.

However, one needs to be more careful if one is dealing with overlapping sets of particles, such as
for example in the case of proton-charge correlations, $\Cov{p,Q}$. 
The protons do carry charge, thus producing a self-correlation term in $\Cov{p,Q}$.
Furthermore, the selection of protons from all the measured charges is typically associated with additional efficiency losses, as one may have to use an additional detector to identify protons.
These effects require a separate efficiency correction treatment.

\section{Efficiency corrections for overlapping sets of particles}
\label{sec:overlap}

There are many off-diagonal cumulants of interest which involve overlapping sets of particles. 
For example, in addition to the aforementioned proton-charge correlations, which may serve as a proxy for baryon number-charge correlations, kaon-charge correlations as proxy for strangeness-charge are also
being studied \cite{Adam:2019xmk}. In the following we will derive the efficiency corrections for
the case of proton-charge correlations noting that the resulting formulas do also apply in other, similar situations, such as kaon-charge correlations etc.

Typically one studies correlations of net numbers, for instance the correlation $\Cov{N_{net-p},Q}$ of the net-proton number with the net charge, $Q=Q^{+}-Q^{-}$.
In order to apply efficiency corrections it is better
to consider the individual terms contributing to  $\Cov{N_{net-p},Q}$
\begin{align}
 \Cov{N_{net-p},Q}=\Cov{N_{p},Q^{+}} + \Cov{N_{\bar{p}},Q^{-}}- \Cov{N_{p},Q^{-}} - \Cov{N_{\bar{p}},Q^{+}}
  \label{eq:Kpq-net}
\end{align}
The last two terms involve only non-overlapping sets of particles, thus the efficiency corrections for
those follow Eq.~\eqref{eq:f_ij_standard} with the appropriate efficiencies.
The first two terms, on the other hand, involve overlapping sets of particles, namely the correlation between numbers of protons and positive charges in the first term, and between anti-protons and negative charges in the second term.

Let us focus on the case of protons and positive charges to derive the necessary efficiency correction formulas. 
These will then straightforwardly apply to other similar cases such as kaon-charge or pion-charge correlations.
To be more specific, we will discuss what we believe is a common scenario in the experiment, where first the sign of the charge is determined e.g. by the bending of tracks in a magnetic field and then the protons are selected from these tracks by an additional particle identification, for instance via a time-of-flight detector.
Let us assume that the efficiencies for
charge and proton identification both follow a binomial distribution. 
We denote
the efficiency for charge
identification by $\eq$  and the efficiency associated with proton identification by $\et$ so
that the total proton detection efficiency is $\ep=\eq \et$.

It is best to separate all positive charges into protons, $N_{p}$ and all other positive charges,
which we denote by $\qh$, so that the total positive charge is given by $\qp=N_{p}+\qh$.
Next, let us denote the {\em true} probability to have $N_{p}$ protons and $\qh$ charges other than
protons by $P(\Np,\qh)$. The {\em true} probability to have $\Np$ protons and $\qp$
positive charges is then $P\left( \Np,\qp  \right) = \sum_{\qh} P\left(\Np,\qh \right)
\delta_{\qp,\Np+\qh} $. The efficiency for charge identification affects both the protons and the other charges, so that the distribution
$p_{\eq}(\npt,\qpm)$ to have $\npt$ protons and $\qpm$ positive charges  {\em after} charge
identification but {\em before} proton identification is
\begin{align}
p_{\eq}\left(\npt,\qpm \right) &= \sum_{\qhm,\qh,\Np}\delta_{\qpm,\qhm+\npt}
  B(\npt,\Np;\eq)B(\qhm,\qh;\eq)P(\Np,\qh) \non
  &= \sum_{\qh,\Np} B(\npt,\Np;\eq)B(\qpm - \npt,\qh;\eq)P(\Np,\qh). 
\label{eq:tpc_only}
\end{align}
The inclusive  probability to measure $\qpm$ charges is then
\begin{align}
  p_{\eq}(\qpm) = \sum_{\npt}p_{\eq}(\npt,\qpm)=\sum_{\qh,\Np}B(\qpm,\Np+\qh;\eq) P(\Np,\qh) 
\end{align}
as it should. Of course we still need to account for the efficiency associated with identifying
proton, $\et$, by folding with the (binomial) probability distribution $B(\npm,\npt,\et)$ for proton
identification. Thus the probability $p_{data}\left( \npm,\qpm \right)$ to measure $\npm$ identified protons and $\qpm$ charges is
given by  
\begin{align}
  p_{data}\left( \npm,\qpm \right) &= \sum_{\npt}B(\npm,\npt;\et) p_{\eq}(\npt,\qpm)
  \non
  &= \sum_{\npt,\qh,\Np}B(\npm,\npt;\et)B(\npt,\Np;\eq)B(\qpm-\npt,\qh;\eq)P(\Np,\qh)
\label{}
\end{align}
Given $p_{data}\left( \npm,\qpm \right)$  the factorial moment generating function is
\begin{align}
  g(\zp,\zq) &= \sum_{\npm,\qhm} \zp^{\npm}\zq^{\qpm}p_{data}\left( \npm,\qpm \right) \non
  & = \sum_{\npm,\qhm,\npt,\qh,\Np}
    \zp^{\npm} \zq^{\qpm}B(\npm,\npt;\et)B(\npt,\Np;\eq)B(\qpm-\npt,\qh;\eq)P(\Np,\qh) \non
   & = \sum_{\Np,\qh}\left[1-\eq + \eq \zq(1-\et + \et \zp) \right]^{\Np}\left( 1-\eq+\zq\eq
     \right)^{\qh}P(\Np,\qh)
  \label{eq:generate_func}
\end{align}
where we used again that $\sum_{n=0}^{N}z^{n}B(n,N;\epsilon)=(1-\epsilon + z \epsilon)^{N}$.
The factorial moments of the {\em measured } distribution, $\ff{i,j}{\npm,\qpm}$ are then easily
obtained\footnote{See Appendix for a more elegant and efficient method to calculate the measured factorial moments.}. 
For example the first factorial moments of the  {\em measured} proton and positive charge
distributions are given by
\begin{align}
  \ff{1,0}{\npm,\qpm}&=\ave{\npm}=\frac{d}{d\zp } \left. g(\zp,\zq)\right|_{\zp=1,\zq=1} =
                      \sum_{\Np,\qh} \eq \et \Np  P(\Np,\qh) =\ep \ave{\Np} = \ep \FF{1,0}{\Np,Q^+} \non
  \ff{0,1}{\npm,\qpm}&=\ave{\qpm}=\frac{d}{d\zq } \left. g(\zp,\zq)\right|_{\zp=1,\zq=1} =
                      \sum_{\Np,\qh} \eq (\Np  + \qh) P(\Np,\qh) = \eq \ave{\qp} =\eq \FF{0,1}{\Np,Q^+}
 \label{eq:f1}
\end{align}
Here we used $\ep=\et \eq$. Similarly, for the diagonal second order factorial moments one finds, after some algebra,
\begin{align}
 \ff{2,0}{\npm,\qpm}&=\ave{\npm (\npm-1)}=\frac{d^{2}}{d\zp^{2}} \left. g(\zp,\zq)\right|_{\zp=1,\zq=1} =
                      \sum_{\Np,\qh} \eq^2 \et^2 \, \Np \left( \Np - 1 \right) 
                     P(\Np,\qh) \non
                     & =  \ep^{2} \ave{\Np(\Np-1)} = \ep^{2} \FF{2,0}{\Np,Q^+} \non
 \ff{0,2}{\npm,\qpm}&=\ave{\qpm (\qpm-1)}=\frac{d^{2}}{d\zq^{2}} \left. g(\zp,\zq)\right|_{\zp=1,\zq=1} =
                      \sum_{\Np,\qh} \eq^{2} \,
                      (\Np + \qh) \, (\Np + \qh - 1) \,
                     P(\Np,\qh) \non
                     &=\eq^{2} \ave{\qp(\qp-1)} =\eq^{2}\FF{0,2}{\Np,Q^+}.
  \label{eq:f2}
\end{align}
Thus we have for the second proton number moment
\begin{align}
\ave{ \npm^{2}} = \ff{2,0}{\npm,\qpm} - \ff{1,0}{\npm,\qpm} =\ep^{2} \FF{2,0}{\Np,\qp} + \ep
  \FF{1,0}{\Np,\qp} = \ep^{2} \left( \ave{\Np^{2}} -\ave{\Np} \right)  + \ep \ave{\Np}~.
\end{align}
The expression for $\ave{{\qpm}^{2}}$ is analogous.

One can see that the measured and true factorial moments involving only protons or only charges follow the standard relation for a binomial efficiency distribution, Eq.~\eqref{eq:f_ij_standard}. 
However, this is no longer the case for the mixed factorial moments involving both charges and protons. 
The measured mixed factorial moment, $\ff{1,1}{\npm,\qpm}$, is given by
\begin{align}
  \ff{1,1}{\npm,\qpm}& =\frac{d^{2}}{d\zp \, d\zq} \left. g(\zp,\zq)\right|_{\zp=1,\zq=1}
                     = \sum_{\Np,\qh} \eq \et \Np 
                     \left[1 + \eq \left( \Np + \qh - 1
                     \right) \right] P(\Np,\qh) \non
                   & = \ep \ave{\Np} + \ep \eq \ave{ \Np \qp - \Np } = \ep\eq \left[ \FF{1,1}{\Np,\qp} +
                     \frac{1-\eq}{\eq} \ave{\Np} \right]~.
  \label{eq:f11}
\end{align}
 We see that the ``standard'' relation, Eq.~\eqref{eq:f_ij_standard}, does not hold anymore. Instead,
expressing the true factorial moment, $\FF{1,1}{\Np,\qp}$, in terms of measured quantities we get,
using $\ave{\Np} = \frac{\ave{\npm}}{\ep}$~[Eq.~\eqref{eq:f1}],
\begin{align}
 \FF{1,1}{\Np,\qp}& = \frac{\ff{1,1}{\npm,\qpm}}{\ep\eq} - \ave{\npm} \frac{1-\eq}{\ep\eq}~, \qquad P \subseteq Q^+.
  \label{eq:F11_correct}
\end{align}
The ``standard'' correction~\eqref{eq:f_ij_standard}, on the other hand, would have only given rise to the first term in this expression. 
Since $\eq<1$, the second term is negative and, therefore, applying the ``standard'' correction would overestimate the true off-diagonal factorial cumulant. 
For the co-variance between protons and positive charges
we then get
\begin{align}
\Cov{N_{p},Q^{+}}=\ave{\delta \Np \delta \qp} &= \FF{1,1}{\Np,\qp} - \FF{1,0}{\Np,\qp}
                                                    \FF{0,1}{\Np,\qp} \non
  &= \frac{1}{\ep\eq} \left( \ff{1,1}{\npm,\qpm} - \ff{1,0}{\npm,\qpm}  \ff{0,1}{\npm,\qpm} \right)
    -\ave{\npm} \frac{1-\eq}{\ep\eq} \non
  &= \frac{1}{\ep\eq} \cov{\npm,\qpm} -\ave{\npm} \frac{1-\eq}{\ep\eq}, \qquad P \subseteq Q^+.
  \label{eq:covariance_correct}
\end{align}
where we have the same extra correction term as compared to the ``standard'' method. 
This term vanishes in the limit $\eq \to 1$.

The same arguments apply also for the co-variance of anti-protons and negative charges $\ave{\delta N_{\bar p} \, \delta Q^{-}}$, the efficiency correction being given by Eq.~\eqref{eq:covariance_correct}.
On the other hand, as the co-variances between protons and negative charges, $\ave{\delta \Np \delta Q^{-}}$, and anti-protons and positive charges,  $\ave{\delta N_{\bar{p}} \delta \qp}$ do not involve overlapping sets of particles, the standard formulas for the efficiency corrections apply, for example, 
\begin{align}
\Cov{\Np,Q^{-}}=\ave{\delta \Np \delta Q^{-}} = \frac{1}{\ep \eqb} \left( \ff{1,1}{\npm,q^{-}} -
  \ff{1,0}{\npm,q^{-}} \ff{0,1}{\npm,q^{-}} \right) =  \frac{1}{\ep \eqb} \ave{\delta
  \npm \delta q^{-}}
  \label{}
\end{align}
and analogous for the $\ave{\delta N_{\bar{p}} \delta \qp}$. Here $\eqb$ denotes the efficiency for
detecting negative charges. Therefore, the co-variance between net-protons and net-charges,
$\Cov{N_{net-p},Q}$~[Eq.~\eqref{eq:Kpq-net}], is given by
\begin{align}
  \Cov{N_{net-p},Q} &= \frac{1}{\ep\eq} \cov{\npm,\qpm} -\ave{\npm} \frac{1-\eq}{\ep\eq}
                       - \frac{1}{\ep\eqb} \cov{\npm,\qmm} \non
    & + \frac{1}{\epb\eqb} \cov{\nbm,\qmm} - \ave{\nbm} \frac{1-\eqb}{\epb\eqb} 
      -  \frac{1}{\epb\eq} \cov{\nbm,\qpm} \non
   & \underset{\ep=\epb,\eq=\eqb}{=}  \frac{1}{\ep\eq} \cov{n_{net-p},q} - \ave{\npm + \nbm} \frac{1-\eq}{\ep\eq}  
\end{align}
where in the last line we assumed identical efficiencies for particles and anti-particles.

In the Appendix we describe how to derive the efficiency corrections for higher-order factorial moments.
A Mathematica notebook that allows to express any desired factorial moment $F_{N_p,Q^+}^{i,j}$ in terms of the measured ones is available via~\cite{effcorr:github}.

\section{Local efficiency corrections}

Our considerations in the previous section are based on the assumption of constant efficiencies.
In a real experiment, however, the efficiency is usually not constant but depends on the momentum of a particle.
For this reason the phase space is typically partitioned into momentum bins, each having its own value of the efficiency parameter.
The local efficiency corrections for the case of a binomial detector response in each bin have been worked out in Refs.~\cite{Bzdak:2013pha,Nonaka:2017kko} for the case of non-overlapping sets of particles.
Here we extend these considerations to cover the case of overlapping sets of particles.
As in Sec.~\ref{sec:overlap}, we shall consider the off-diagonal cumulants of protons and charged particles as a concrete example.

Let us assume an arbitrary partition of the phase space into bins. 
Following  Ref.~\cite{Bzdak:2013pha}, we introduce a variable $x$ to enumerate the bins.
The numbers of (anti)protons and positively~(negatively) charged particles are obtained by summing over all the bins:
\eqn{\label{eq:localN}
N_{p(\bar{p})} & = \sum_x N_{p(\bar{p})} (x), \non
Q^\pm & = \sum_x Q^\pm (x), \non
n_{p(\bar{p})} & = \sum_x n_{p(\bar{p})} (x), \non
q^\pm & = \sum_x q^\pm (x).
}
Here $N_{p(\bar{p})} (x)$ and $Q^\pm (x)$ correspond to the numbers of (anti)protons and positive~(negative) charges in a phase-space bin $x$.
As before, the lowercase $n_{p(\bar{p})} (x)$ and $q^\pm (x)$ correspond to the numbers of measured particles.

Consider now the factorial moments involving protons and positive charges.
The first-order proton number moments read
\eqn{
F_{N_p,Q^+}^{1,0} & = \ave{N_p} = \sum_{x} \, \ave{N_p(x)}, \non
f_{n_p,q^+}^{1,0} & = \ave{n_p} = \sum_{x} \, \ave{n_p(x)}.
}
As the binomial efficiency corrections in different phase space-bins are independent, one has $\ave{n_p(x)} = \ep(x) \ave{N_p(x)}$ and, thus,
\eqn{
F_{N_p,Q^+}^{1,0} & = \ave{N_p} = \sum_{x} \, \frac{\ave{n_p(x)}}{\epsilon_p(x)},
}
and, analogously for positive charges,
\eqn{
F_{N_p,Q^+}^{0,1} & = \ave{Q^+} = \sum_{x} \, \frac{\ave{q^+(x)}}{\epsilon_q(x)}.
}
The second-order diagonal proton number factorial moments are
\eqn{\label{eq:loctwo}
F_{N_p,Q^+}^{2,0} & = \ave{N_p (N_p - 1)} = \sum_{x_1,x_2} \, \ave{N_p(x_1) [N_p(x_2) - \delta_{x_1,x_2}] }, \non
f_{n_p,q^+}^{2,0} & = \ave{n_p (n_p - 1)} = \sum_{x_1,x_2} \, \ave{n_p(x_1) [n_p(x_2) - \delta_{x_1,x_2}] } .
}
The binomial efficiency correction is applied independently for all pairs of bins in~\eqref{eq:loctwo}, as per Eq.~\eqref{eq:f2}, therefore,
\eqn{
F_{N_p,Q^+}^{2,0} & = \ave{N_p (N_p - 1)} 
= \sum_{x_1,x_2} \, \frac{\ave{n_p(x_1) [n_p(x_2) - \delta_{x_1,x_2}] }}{\epsilon_p(x_1) \, \epsilon_p(x_2)},
}
and, analogously,
\eqn{
F_{N_p,Q^+}^{0,2} & = \ave{Q^+ (Q^+ - 1)} 
= \sum_{x_1,x_2} \, \frac{\ave{q^+(x_1) [q^+(x_2) - \delta_{x_1,x_2}] }}{\epsilon_q(x_1) \, \epsilon_q(x_2)}~.
}
These results for the diagonal factorial moments are the same as obtained in Ref. \cite{Bzdak:2013pha}.
Consider now the mixed factorial moments
\eqn{
F_{N_p,Q^+}^{1,1} & = \ave{N_p \, Q^+} = \sum_{x_1,x_2} \, \ave{N_p(x_1) Q^+(x_2) }, \non
f_{n_p,q^+}^{1,1} & = \ave{n_p \, q^+} = \sum_{x_1,x_2} \, \ave{n_p(x_1) q^+(x_2) }~.
\label{eq:localmixed}
}
The efficiency correction proceeds independently for all pairs $(x_1,x_2)$ of bins, as advocated above.
The terms with $x_1 \neq x_2$ correspond to protons and positive charges from different phase-space bins, which, therefore, correspond to non-overlapping particles.
For these terms the ``standard'' correction~[Eq.~\eqref{eq:f_ij_standard}] applies.
However, the terms with $x_1 = x_2$ in Eq.~\eqref{eq:localmixed} correspond to mixed factorial moments involving overlapping sets of particles: protons and positive charges in the same phase-space bin.
This means that the generalized correction~\eqref{eq:F11_correct} should be used in these cases.
Combining the corrections for the $x_1 \neq x_2$ and $x_1 = x_2$ terms together, we arrive at
\eqn{
F_{N_p,Q^+}^{1,1} & = \ave{N_p \, Q^+} = \sum_{x_1,x_2} \, \frac{\ave{n_p(x_1) q^+(x_2) }}{\epsilon_p(x_1) \, \epsilon_q(x_2)} - \sum_{x} \, \ave{n_p(x)} \, \frac{1-\epsilon_q(x)}{\epsilon_p(x) \, \epsilon_q(x)}~.
}

The relation between the true co-variance between protons and positive charges and the measured moments reads
\begin{align}
\Cov{N_{p},Q^{+}}=\ave{\delta \Np \, \delta \qp} &= \FF{1,1}{\Np,\qp} - \FF{1,0}{\Np,\qp}
                                                    \FF{0,1}{\Np,\qp} \non
  &= \sum_{x_1,x_2} \, \frac{\cov{n_p(x_1) q^+(x_2) }}{\epsilon_p(x_1) \, \epsilon_q(x_2)} - \sum_{x} \, \ave{n_p(x)} \, \frac{1-\epsilon_q(x)}{\epsilon_p(x) \, \epsilon_q(x)}~.
  \label{eq:covariance_local}
\end{align}

The expression for the co-variance between antiprotons and negative charges is analogous to~\eqref{eq:covariance_local}.
For the co-variances between protons and negative charges, as well as between antiprotons and positive charges, one does not encounter overlapping sets of particles, thus the second term in the r.h.s. of Eq.~\eqref{eq:covariance_local} does not appear.
For completeness, we list here the results for all the remaining proton-charge covariances:
\eqn{\label{eq:covariance_local2}
\Cov{N_{\bar{p}},Q^{-}} &= \ave{\delta \Npb \, \delta \qm} = \sum_{x_1,x_2} \, \frac{\cov{n_{\bar{p}}(x_1) q^-(x_2) }}{\epb(x_1) \, \eqb(x_2)} - \sum_{x} \, \ave{n_{\bar{p}}(x)} \, \frac{1-\eqb(x)}{\epb(x) \, \eqb(x)}~, \\
\label{eq:covariance_local3}
\Cov{N_p,Q^{-}} & =\ave{\delta \Np \, \delta \qm} = \sum_{x_1,x_2} \, \frac{\cov{n_{p}(x_1) q^-(x_2) }}{\ep(x_1) \, \eqb(x_2)}, \\
\label{eq:covariance_local4}
\Cov{N_{\bar{p}},Q^{+}} & = \ave{\delta \Npb \, \delta \qm} = \sum_{x_1,x_2} \, \frac{\cov{n_{\bar{p}}(x_1) q^+(x_2) }}{\epb(x_1) \, \eq(x_2)}~.
}
The co-variance between net protons and net charges is then given by Eq.~\eqref{eq:Kpq-net}.
Equations~\eqref{eq:covariance_local}-\eqref{eq:covariance_local4} reduce to the results of Sec.~\ref{sec:overlap} for the case of uniform efficiencies.

The local efficiency correction procedure is relevant for the measurements of the off-diagonal susceptibilities that are being performed by the STAR experiment. 
As detailed in Ref.~\cite{Adam:2019xmk}, the measurements are done in a range $0.4 < p_T < 1.6$~GeV/$c$.
This range is split into two bins, $0.4 < p_T < 0.8$~GeV/$c$ and $0.8 < p_T < 1.6$~GeV/$c$, which differ by the proton identification procedure.
Protons from the bin $0.4 < p_T < 0.8$~GeV/$c$ are identified by applying cuts on the track parameters extracted using the Time Projection Chamber~(TPC) and do not involve any additional measurements.
This means that the same detector~(TPC) is used for both the charge selection and particle identification for this $p_T$ bin.
To identify protons from the second bin, $0.8 < p_T < 1.6$~GeV/$c$, an additional measurement using a time-of-flight detector is performed.
The efficiency correction for these STAR measurements can thus be performed by applying the formalism presented in this section for the case of two bins.
The proton efficiencies in each of the bins will be different, reflecting the difference in the procedure to identify protons described above.

\section{Examples}

\subsection{Poisson distributed particles}

To illustrate the above findings let us consider the case where all particle multiplicities are independent and follow Poisson distributions. 
Let us consider again the case of protons and positive charges, and uniform binomial efficiencies, i.e. a single phase-space bin.
With $\qp=\Np+\qh$ the true co-variance between protons and positively charged
particles is given by
\begin{align}
\Cov{\Np,\qp}=\ave{\delta \Np \, \delta \qp} =  \ave{\delta \Np \,\delta \qh} +  \ave{\left(\delta  \Np \right)^{2}}
  = \ave{\left( \delta \Np \right)^{2} } \underset{\rm Poisson}{=} \ave{\Np}
  \label{eq:covariance_poisson}
\end{align}
Here we used the fact that  $\Cov{\Np,\qh}=  \ave{\delta \Np \, \delta \qh} =0$ for independently distributed $\Np$ and $\qh$. 
The only source of correlations between numbers of protons and positive charges is the proton self-correlation.

Next let us calculate the same co-variance for the measured particles, $\cov{\npm,\qpm} =
\ave{\left( \delta \npm \delta \qpm \right) } = \ff{1,1}{\npm,\qpm} - \ave{\npm}\ave{\qpm}$. Since the particles are distributed independently, the   
probability distribution $P(\Np,\qh)$ factorizes into that for protons and that for all other
positively charged particles, 
\begin{align}
P(\Np,\qh) = P_{p}\left(\Np,\ave{\Np}\right) \, P_{p}\left(\qh,\ave{\qh}\right)
  \label{}
\end{align}
with $P_{p}(N,\Lambda)=\exp(-\Lambda) \Lambda^{N}/N!$ denoting a Poisson distribution with mean
$\ave{N}=\Lambda$. With $\sum_{N=0}^{\infty}z^{N} P_{p}(N,\Lambda) = \exp(\Lambda(z-1))$ the factorial moment
generating function, Eq.~\eqref{eq:generate_func}, is readily evaluated
\begin{align}
  g(\zp,\zq) &= \exp\left[\ave{\Np} \ep \zq(\zp-1) \right] \exp\left[\ave{\qp} \eq (\zq-1)
               \right] 
  \label{}
\end{align}
where we have used $\ave{\qp}=\ave{\qh}+\ave{\Np}$. Given the generating function, the factorial
moment $\ff{1,1}{\npm,\qpm}$ is
\begin{align}
  \ff{1,1}{\npm,\qpm}& =\frac{\partial^{2}}{\partial \zp \, \partial \zq} \left. g(\zp,\zq)\right|_{\zp=1,\zq=1}\non
                   & = \ep \eq \ave{\Np} \ave{\qp} + \ep \ave{\Np}.   
  \label{eq:f11_poisson}
\end{align}
With $\ave{\npm}=\ep \ave{\Np} $ and $\ave{\qpm} =\eq \ave{\qp}$ we get for the measured co-variance
\begin{align}
\cov{\npm,\qpm} = \ff{1,1}{\npm,\qpm} - \ave{\npm}\ave{\qpm} =  \ep \ave{\Np}=\ave{\npm}
  \label{}
\end{align}
which is the expected result for Poisson distributed particles since the binomial efficiency
correction result again in Poisson distributed measured particles. And, applying the efficiency
correction, Eq.~\eqref{eq:covariance_correct}, we recover the result for the true distribution,
Eq.~\eqref{eq:covariance_poisson}
\begin{align}
  \Cov{\Np,\qp} = \frac{\cov{\npm,\qp}}{\ep\eq} - \ave{\npm} \frac{1-\eq}{\ep\eq} 
                  =  \frac{\ave{\Np}}{\eq} - \ave{\npm} \frac{1-\eq}{\ep\eq} = \frac{\ave{\npm}}{\eq} = \ave{\Np}.
  \label{eq:f11_correct_poisson}
\end{align}
However, if we were to apply the ``standard'' efficiency correction, Eq.~\eqref{eq:f_ij_standard}, which applies only to
non-overlapping sets of particles, we would get
\begin{align}
  \frac{\cov{\npm,\qpm}}{\ep\eq} = \frac{\ave{\Np}}{\eq} =  \frac{\Cov{\Np,\qp}}{\eq}
  \label{eq:covariance_wrong}
\end{align}
which for $\eq < 1$ would (incorrectly) 
suggest the presence of extra positive correlations in addition to the proton self-correlation.

\subsection{Monte Carlo simulation of a case with a non-trivial proton-charge correlation}
\label{sec:MC}

Finally, we shall test the developed efficiency correction for a more general case when non-trivial proton-charge correlations are present.
Here we perform a Monte Carlo simulation, which resembles more closely the actual procedure done in experiment.
As in the previous example, we only look here at protons and positive charges.

We assume the following setup.
Each event is characterized by three non-negative integer numbers, $N_1$, $N_2$, and $N_3$. 
These three numbers are all independent and distributed in accordance with a Poisson distribution, i.e. $P(N_i) = P_p (N_i, \ave{N_i})$ for $i = 1,2,3$.
The values of $N_1$, $N_2$, $N_3$ define the numbers of protons $N_p$ and other positive charges $\qh$ in a given event as follows:
\eqn{
\Np & = N_1 + N_2, \\
\qh & = N_2 + N_3.
}
The fact that $N_2$ contributes to both the number of protons and that of other positive charges generates a non-trivial proton-charge correlation.\footnote{One example for such a correlation would be the decay of the $\Delta^{++}$ into a proton and a positively charged pion, thus contributing to both  $\Np$ and $\qh$. }
Evaluating $\Cov{\Np,\qp}$ explicitly yields
\eqn{
\Cov{\Np,\qp} & = \ave{\delta \Np \, \delta \qp} = \ave{\delta \Np^2 } + \ave{\delta \Np \, \delta \qh} \non
& = \ave{\delta N_1^2 } + 2 \ave{\delta N_2^2 } + 3 \ave{\delta N_1 \delta N_2 } + \ave{\delta N_1 \delta N_3 } + \ave{\delta N_2 \delta N_3 } \non
& = \ave{N_1} + 2 \ave{N_2} \non
& = \ave{N_p} + \ave{N_2}.
}
Normalizing $\Cov{\Np,\qp}$ by the mean number of protons, one can explicitly see the additional proton-charge correlation in excess of the proton self-correlation:
\eqn{\label{eq:cov_mc_expected}
\frac{\Cov{\Np,\qp}}{\ave{\Np}} & = 1 + \frac{\ave{N_2}}{\ave{N_1} + \ave{N_2}}~.
}

The Monte Carlo simulation procedure is the following:
\begin{enumerate}
    \item For each event first the numbers $N_1$, $N_2$, and $N_3$ are sampled from the three independent Poisson distributions. The numbers of protons and other positive charges are then evaluated as $N_p = N_1 + N_2$ and $\qh = N_2 + N_3$.
    \item The charge identification efficiency is simulated by applying a binomial filter with efficiency $\eq$ to both the number of protons $N_p$ and other positive charges $\qh$. This gives the number of charge-identified protons $\tilde{n}_p$ and other positive charges $\qhm$.
    The total measured charge in the given event is registered as $q^+ = \tilde{n}_p + \qhm$.
    \item The proton identification efficiency is simulated by applying a additional binomial filter with efficiency $\et \equiv \ep / \eq$ to the number of the charge-identified protons $\tilde{n}_p$.
    This gives the number of identified protons $n_p$ in the given event.
    \item The factorial moments $\ff{i,k}{\npm,\qpm}$ of identified protons and positive charges are evaluated as statistical averages over all the simulated events.
\end{enumerate}

Here we present results of the simulations for parameter values $\ave{N_1} = 90$, $\ave{N_2} = 10$, and $\ave{N_3} = 200$.
We fix the proton identification efficiency at $\et = 0.7$ but vary the charge identification efficiency in a range $0.5 < \eq < 1$ in steps of 0.05.
For each value of $\eq$ we generate one million events in accordance with the procedure described above.
The true proton-charge co-variance $\Cov{\Np,\qp}$ is reconstructed using the factorial moments $\ff{i,k}{\npm,\qpm}$ of measured particle numbers, calculated as averages over the sampled events, and the efficiency correction via Eq.~\eqref{eq:covariance_correct}.
The results are compared with the ``standard'' efficiency correction~[Eq.~\eqref{eq:cov_ij_standard}].

\begin{figure}[t]
  \centering
  \includegraphics[width=.64\textwidth]{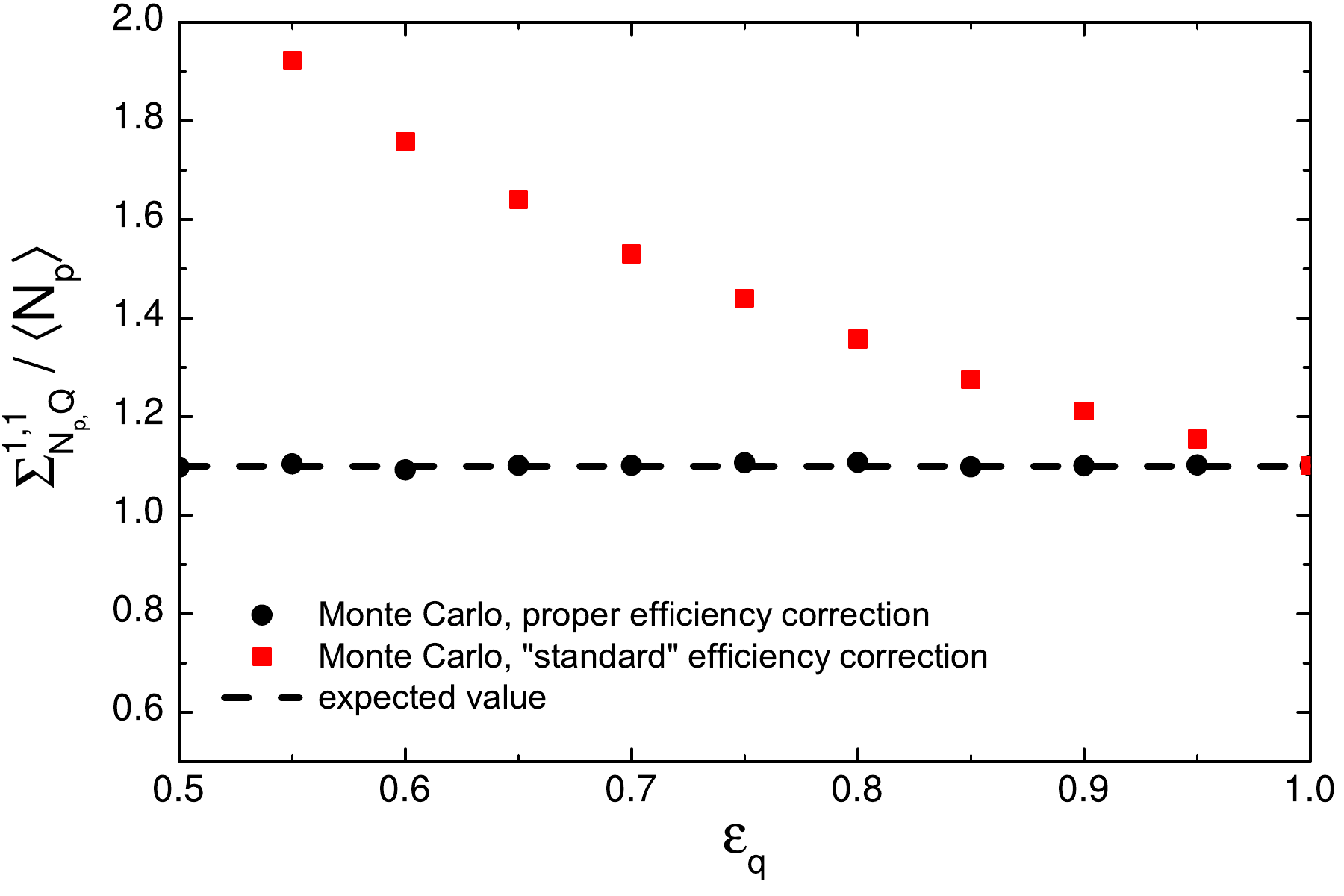}
  \caption{
  The reconstructed values of the scaled proton-charge correlator $\Cov{\Np,\qp}/\ave{\Np}$ for various values of the charge identification efficiency $\eq$ from Monte Carlo simulations of a toy model described in Sec.~\ref{sec:MC}.
  For each value of $\eq$ one million events was generated.
  The black circles depict the results obtained by applying the proper efficiency correction via Eq.~\eqref{eq:covariance_correct}.
  The red squares correspond to the ``standard'' efficiency correction~\eqref{eq:cov_ij_standard}, which is applicable only for the case of non-overlapping particles.
  The horizontal dashed line corresponds to the true value of $\Cov{\Np,\qp}/\ave{\Np} = 1.1$.
  }
  \label{fig:effcorMC}
\end{figure}

Figure~\ref{fig:effcorMC} depicts the reconstructed values of the scaled proton-charge correlator $\Cov{\Np,\qp}/\ave{\Np}$ for various values of $\eq$.
The black symbols correspond to the results obtained using the proper efficiency corrections given by Eq.~\eqref{eq:covariance_correct}.
For all the $\eq$ values the efficiency corrected Monte Carlo results are consistent with the true value of $\Cov{\Np,\qp}/\ave{\Np} = 1.1$, as given by Eq.~\eqref{eq:cov_mc_expected}.
This agreement validates the efficiency correction derived in this work.
On the other hand, the ``standard'' efficiency correction, the result of which for $\Cov{\Np,\qp}/\ave{\Np}$ is shown in Fig.~\ref{fig:effcorMC} by the red symbols, systematically overestimates the true value of proton-charge correlations.
The error is larger for smaller values of $\eq$ and only disappears in the limit of perfect charge identification, $\eq \to 1$.
The realistic range for $\eq$ in heavy-ion collisions, however, is of order $\eq = 0.6-0.8$~\cite{Abelev:2012pv,Adam:2019xmk}.
Using the ``standard'' efficiency correction for such values of $\eq$ leads to an overestimation of $\Cov{\Np,\qp}$ by as much as 20-50\%.
This underscores the importance of taking into account the subtleties associated with the efficiency corrections for overlapping sets of particles discussed in the present note.

\section{Discussion and Summary}
\begin{itemize}
   \item We note that the difference between the true correlation $\FF{1,1}{\Np,\qp}=\ave{\Np \, \qp}$
    and that obtained from the measured quantities via the ``standard'' efficiency correction, Eq.~\eqref{eq:f11},
    \begin{align}
      \FF{1,1}{\Np,\qp} - \frac{\ff{1,1}{\npm,\qpm}}{\ep \eq} = - \ave{\npm} \frac{1-\eq}{\ep\eq}
      \label{}
    \end{align}
    vanishes in the limit of perfect charge detection, $\eq \to 1$. This is not surprising, since in
    this case one only needs to correct for the proton detection efficiency and the corrections for
    protons only do agree with the standard procedure, as shown above.
  \item It may be instructive to consider the measured correlation between protons and all other
    positive charged particles, as this corresponds to a correlation of non-overlapping {\em
      measured} particles. Using Eqs.~\eqref{eq:f1},~\eqref{eq:f2}, and \eqref{eq:F11_correct}
    one gets
    \begin{align}
      \ave{\npm \qhm} =\ave{\npm \qpm} - \ave{\npm^{2}} = \ff{1,1}{\npm,\qpm} - \ave{\npm^{2}}  =
      \ep \eq \ave{\Np \qh} + \ep(\eq-\ep) \left( \ave{\Np^{2}} - \ave{\Np} \right) 
      \label{}
    \end{align}
    Even in this case, of seemingly non-overlapping particles, the ``standard'' correction does not
    work in general, since 
    \begin{align}
      \ave{\Np \qh} - \frac{\ave{\npm \qhm}}{ \ep \eq } = \frac{(\ep-\eq)}{\eq} \left( \ave{\Np^{2}}
      - \ave{\Np} \right) \ne 0.
    \end{align}
    Only if we have perfect proton identification, i.e. $\et = \ep/\eq = 1$ does the ``standard''
    correction work. This is easy to understand. With perfect proton identification we remove
    all protons from the measured charge when we calculate $\qhm = q^+-\npm$. Otherwise, $q^+$ always contains protons, which are identified as charges but not as protons. This implies that the finite detection efficiency induces artificial correlations between the measured
    protons, $\npm$, and the measured other positive charge, $\qhm$. Indeed, calculating the co-variance between
    protons and other charges, $\cov{\npm,\qhm}$ in the case where  the true distribution is uncorrelated,
    $\Cov{\Np,\qh}=0$, one finds
    \begin{align}
      \cov{\npm,\qhm} = \ep (\eq-\ep) \left[ \ave{(\delta \Np)^{2}} -\ave{\Np} \right]
    \end{align}
    which vanishes only in the special case of Poisson distributed protons. 
  
  \item Efficiency corrections for mixed cumulants of (possibly) overlapping sets of particles have been discussed in Ref.~\cite{Nonaka:2017kko}, also based on the binomial efficiency model. 
  The results obtained in that paper apply in the case when one can separately measure the multiplicity for every particle species that contributes to the sets under consideration.
  This is different from the present work: due to an imperfect proton identification efficiency, $\et < 1$, we cannot reconstruct the number $\hat{q}^+$ of positive charges other than protons that contribute to the measured number $q^+$ of all positive charges in a given event~(see the discussion in the previous bullet point).
  For this reason, the efficiency correction formulas of Ref.~\cite{Nonaka:2017kko} are different from the present work, and only coincide with our result in the limiting case of perfect proton identification, $\et = \ep/\eq = 1$.
  
  \item The presented procedure can be extended to higher order factorial moments and cumulants. As
    detailed in the Appendix, given the factorial cumulant generating function,
    Eq.~\eqref{eq:generate_func}, one can calculate the measured factorial cumulants in terms of the
    true ones and then simply needs to invert these relations in order to obtain expression for the
    true factorial moments in terms of measured quantities.
    
  \item The local efficiency corrections proceed by correcting all the relevant local factorial moments, as originally devised in Ref.~\cite{Bzdak:2013pha}. In contrast to~\cite{Bzdak:2013pha}, however, the generalized efficiency correction must be used for those local factorial moments that involve overlapping particles from the same phase-space bin.
    
  \item While we have mostly concentrated on the specific case of protons and positive charges, the above results do apply to all cases of overlapping particles, such as
  kaon-charge correlators $\ave{\delta N_{K^+} \delta Q^+}$ and $\ave{\delta N_{K^-} \delta Q^-}$, and others, as long as detection of the identified particle, say the proton or kaon, involves the same charge identification process as all other charges. 
  If, on the other hand, one had two {\em distinct} detectors, one to measure all charges, one to measure the protons or kaons without making use of the charge measurement of the other detector, then the standard procedure works. 
  As in this case the identification of a proton does not influence in any way the identification of all the charged particles, including the same proton, and vice versa.
  
  \item We note that proton-charge correlations can be analyzed in an alternative way, namely one would first identify all the individual charged particles, including pions, kaons, and protons. 
  Then, by defining the number of charged particles as the sum of the individual identified charged particles, one can construct the charge-proton correlation from the correlators involving the identified particles.
  As these correlators would always involve non-overlapping sets of particles, one can use the standard procedure to perform the efficiency correction.
  The method introduced in the present paper does have several advantages over this alternative procedure.
  In particular, our method does not involve identification of particles other than protons, thus it suffers smaller efficiency losses and it is less prone to issues involving particle misidentification.
  
  \end{itemize}  

In summary, we have derived the formulas for the efficiency corrections of co-variances involving overlapping sets of particles. 
These formulas apply when the same detector is used for the initial identification of all the particles, as is the case e.g. for the reconstruction of charged tracks in heavy-ion collision experiments, 
and then a subset of these particles, such as protons among all the charged particles, 
is identified via an additional procedure that may involve another detector.
Our main result here is Eq.~\eqref{eq:covariance_correct}, which shows that an extra term arises compared to the case of distinct particles, which would result in apparent larger correlation if not properly taken into account. 
The result has also been generalized for the case of local efficiency corrections~[Eq.~\eqref{eq:covariance_local}], which permit variations in the efficiencies between different phase-space bins.

\section*{Acknowledgments}
\hspace*{\parindent}
We acknowledge fruitful discussions with Arghya Chatterjee and Dmytro Oliinychenko. 
We also thank Masakiyo Kitazawa for constructive comments.
This material is based upon work supported by the U.S. Department of Energy, Office of Science, Office of Nuclear Physics, under contract number 
DE-AC02-05CH11231.
The work of V.V. received support through the Feodor Lynen program of the Alexander von Humboldt foundation.

\appendix
\section{}
\label{sec:appendix_a}
Here we present a more elegant and efficient way of relating the measured factorial moments,
$\ff{i,j}{\npm,\qpm}$ with those of the true distribution, $\FF{i,j}{\Np,\qp}$. 
We again restrict ourselves to the specific case of protons and positive charges noting that the results can be directly translated to other equivalent cases, such as $K^{-}$ and negative charges etc.
Let us start with the factorial moment generating function for the true distribution,
$G(\Zp,\Zq)$. Given the probability function $P(\Np,\qp) = \sum_{\qh} P(\Np,\qh) \delta_{\qp,\qh+\Np}$, the
generating function is given by
\begin{align}
G(\Zp,\Zq) = \sum_{\Np,\qp} \Zp^{\Np} \Zq^{\qp} P(\Np,\qp) =
  \sum_{\Np,\qh}  \Zp^{\Np} \Zq^{\qh+\Np} P(\Np,\qh)
  \label{eq:app:cum_gen_true}
\end{align}
The true factorial moments are 
\begin{align}
\FF{i,j}{\Np,\qp} = \left. \frac{\partial^{i+j}}{\partial \Zp^{i} \partial \Zq^{j}}G(\Zp,\Zq)\right|_{\Zp=\Zq=1}
  \label{}
\end{align}
Comparing with the expression of the generating function for the factorial cumulants of the measured distribution,
$g(\zp,\zq)$~[Eq.~\eqref{eq:generate_func}], we find that $g(\zp,\zq) $ can be expressed in terms of $G(\Zp,\Zq)$
\begin{align}
g(\zp,\zq) = G\left[ \Zp(\zp,\zq),\Zq(\zq) \right]
  \label{}
\end{align}
with
\begin{align}
\Zp(\zp,\zq) &= \frac{1-\eq + \eq \zq(1-\et + \et \zp) }{ 1-\eq+\zq\eq}
               \non
\Zq(\zq) &= 1-\eq+\zq\eq 
  \label{}
\end{align}
Therefore, as we have $\Zp(\zp=1,\zq=1)=\Zq(\zq=1)=1$, the measured factorial moments
\begin{align}
\ff{i,j}{\npm,\qpm} = \left. \frac{\partial^{i+j}}{\partial \zp^{i} \partial
  \zq^{j}}g(\zp,\zq)\right|_{\zp=\zq=1}  = \left. \frac{\partial^{i+j}}{\partial \zp^{i} \partial
  \zq^{j}}G\left( \Zp(\zp,\zq),\Zq(\zq) \right)  \right|_{\zp=\zq=1}
    \label{}
\end{align}
can be easily related to those of the true distribution by applying the chain rule. For
example
\begin{align}
  \ff{1,0}{\npm,\qpm} &= \left. \frac{\partial}{\partial \zp} G\left( \Zp(\zp,\zq),\Zq(\zq) \right)
                       \right|_{\zp=\zq=1}
                       \non
  &= \left. \frac{\partial}{\partial \Zp} G\left( \Zp,\Zq \right) \right|_{\Zp=\Zq=1} \times
  \left. \frac{\partial}{\partial \zp}  \Zp(\zp,\zq) \right|_{\zp=\zq=1} \non
  & = \ep \FF{1,0}{\Np,\qp} \non
  \label{}
\end{align}
where in the last step we used the expression for the full detection efficiency for the protons,
$\ep=\eq \et$.
Many terms in these expressions will vanish as only a few of the derivatives of $\Zp(\zp,\zq)$ and $\Zq(\zq)$ are nonzero:
\begin{align}
  \left. \frac{\partial}{\partial \zq}  \Zq(\zq) \right|_{\zp=\zq=1}&=\eq,
  \non
  \left. \frac{\partial^{n}}{\partial \zq^{n}}  \Zq(\zq) \right|_{\zp=\zq=1}&=0,\quad n>1
  \non
  \left. \frac{\partial}{\partial \zp}  \Zp(\zp,\zq) \right|_{\zp=\zq=1}&= \ep, \non
  \left. \frac{\partial^{n}}{\partial \zp^{n}}  \Zp(\zp,\zq) \right|_{\zp=\zq=1} &= 0, \quad n>1
  \non
   \left. \frac{\partial^{n+1}}{\partial \zp \partial \zq^{n}}  \Zp(\zp,\zq) \right|_{\zp=\zq=1} &=
    (-1)^{n + 1} n!\, \ep (1 - \eq) \eq^{n - 1}. 
  \label{eq:qpp:z_derivatives}
\end{align}
This procedure, which can be easily automatized for high-order factorial moments using tools such as Mathematica, provides the
measured factorial moments expressed in terms of the true factorial moments. 
These can then be
inverted to obtain the relations that express the true factorial moments in terms of the measured ones, 
which, in turn, provide the necessary relations required for efficiency corrections. 
For example, 
the third-order mixed factorial moments 
are given by
\begin{align}
  \FF{2,1}{\Np,\qp} &= \frac{\ff{2,1}{\npm,\qpm}}{\ep^{2}\eq} 
   - \frac{2 (\eq-1)\ff{2,2}{\npm,\qpm}}{\ep^{2}\eq} \non
  \FF{1,2}{\Np,\qp} &= \frac{\ff{1,2}{\npm,\qpm}}{\ep \eq^{2}}
    - \frac{2 (\eq-1)\left( \ff{1,1}{\npm,\qpm} - \ff{1,0}{\npm,\qpm} \right) }{\ep \eq^{2}}
  \label{}
\end{align}

A Mathematica notebook carrying out this task for any desired factorial moment is available via~\cite{effcorr:github}.


\bibliography{off_diagonal_efficiency}

\begin{thebibliography}{18}%
\makeatletter
\providecommand \@ifxundefined [1]{%
 \@ifx{#1\undefined}
}%
\providecommand \@ifnum [1]{%
 \ifnum #1\expandafter \@firstoftwo
 \else \expandafter \@secondoftwo
 \fi
}%
\providecommand \@ifx [1]{%
 \ifx #1\expandafter \@firstoftwo
 \else \expandafter \@secondoftwo
 \fi
}%
\providecommand \natexlab [1]{#1}%
\providecommand \enquote  [1]{``#1''}%
\providecommand \bibnamefont  [1]{#1}%
\providecommand \bibfnamefont [1]{#1}%
\providecommand \citenamefont [1]{#1}%
\providecommand \href@noop [0]{\@secondoftwo}%
\providecommand \href [0]{\begingroup \@sanitize@url \@href}%
\providecommand \@href[1]{\@@startlink{#1}\@@href}%
\providecommand \@@href[1]{\endgroup#1\@@endlink}%
\providecommand \@sanitize@url [0]{\catcode `\\12\catcode `\$12\catcode
  `\&12\catcode `\#12\catcode `\^12\catcode `\_12\catcode `\%12\relax}%
\providecommand \@@startlink[1]{}%
\providecommand \@@endlink[0]{}%
\providecommand \url  [0]{\begingroup\@sanitize@url \@url }%
\providecommand \@url [1]{\endgroup\@href {#1}{\urlprefix }}%
\providecommand \urlprefix  [0]{URL }%
\providecommand \Eprint [0]{\href }%
\providecommand \doibase [0]{http://dx.doi.org/}%
\providecommand \selectlanguage [0]{\@gobble}%
\providecommand \bibinfo  [0]{\@secondoftwo}%
\providecommand \bibfield  [0]{\@secondoftwo}%
\providecommand \translation [1]{[#1]}%
\providecommand \BibitemOpen [0]{}%
\providecommand \bibitemStop [0]{}%
\providecommand \bibitemNoStop [0]{.\EOS\space}%
\providecommand \EOS [0]{\spacefactor3000\relax}%
\providecommand \BibitemShut  [1]{\csname bibitem#1\endcsname}%
\let\auto@bib@innerbib\@empty
\bibitem [{\citenamefont {Bzdak}\ \emph {et~al.}(2020)\citenamefont {Bzdak},
  \citenamefont {Esumi}, \citenamefont {Koch}, \citenamefont {Liao},
  \citenamefont {Stephanov},\ and\ \citenamefont {Xu}}]{Bzdak:2019pkr}%
  \BibitemOpen
  \bibfield  {author} {\bibinfo {author} {\bibfnamefont {A.}~\bibnamefont
  {Bzdak}}, \bibinfo {author} {\bibfnamefont {S.}~\bibnamefont {Esumi}},
  \bibinfo {author} {\bibfnamefont {V.}~\bibnamefont {Koch}}, \bibinfo {author}
  {\bibfnamefont {J.}~\bibnamefont {Liao}}, \bibinfo {author} {\bibfnamefont
  {M.}~\bibnamefont {Stephanov}}, \ and\ \bibinfo {author} {\bibfnamefont
  {N.}~\bibnamefont {Xu}},\ }\href {\doibase 10.1016/j.physrep.2020.01.005}
  {\bibfield  {journal} {\bibinfo  {journal} {Phys. Rept.}\ }\textbf {\bibinfo
  {volume} {853}},\ \bibinfo {pages} {1} (\bibinfo {year} {2020})},\ \Eprint
  {http://arxiv.org/abs/1906.00936} {arXiv:1906.00936 [nucl-th]} \BibitemShut
  {NoStop}%
\bibitem [{\citenamefont {Stephanov}(2009)}]{Stephanov:2008qz}%
  \BibitemOpen
  \bibfield  {author} {\bibinfo {author} {\bibfnamefont {M.}~\bibnamefont
  {Stephanov}},\ }\href {\doibase 10.1103/PhysRevLett.102.032301} {\bibfield
  {journal} {\bibinfo  {journal} {Phys.Rev.Lett.}\ }\textbf {\bibinfo {volume}
  {102}},\ \bibinfo {pages} {032301} (\bibinfo {year} {2009})},\ \Eprint
  {http://arxiv.org/abs/0809.3450} {arXiv:0809.3450 [hep-ph]} \BibitemShut
  {NoStop}%
\bibitem [{\citenamefont {Morita}\ \emph {et~al.}(2013)\citenamefont {Morita},
  \citenamefont {Friman}, \citenamefont {Redlich},\ and\ \citenamefont
  {Skokov}}]{Morita:2013tu}%
  \BibitemOpen
  \bibfield  {author} {\bibinfo {author} {\bibfnamefont {K.}~\bibnamefont
  {Morita}}, \bibinfo {author} {\bibfnamefont {B.}~\bibnamefont {Friman}},
  \bibinfo {author} {\bibfnamefont {K.}~\bibnamefont {Redlich}}, \ and\
  \bibinfo {author} {\bibfnamefont {V.}~\bibnamefont {Skokov}},\ }\href
  {\doibase 10.1103/PhysRevC.88.034903} {\bibfield  {journal} {\bibinfo
  {journal} {Phys. Rev.}\ }\textbf {\bibinfo {volume} {C88}},\ \bibinfo {pages}
  {034903} (\bibinfo {year} {2013})},\ \Eprint {http://arxiv.org/abs/1301.2873}
  {arXiv:1301.2873 [hep-ph]} \BibitemShut {NoStop}%
\bibitem [{\citenamefont {Koch}\ \emph {et~al.}(2005)\citenamefont {Koch},
  \citenamefont {Majumder},\ and\ \citenamefont {Randrup}}]{Koch:2005vg}%
  \BibitemOpen
  \bibfield  {author} {\bibinfo {author} {\bibfnamefont {V.}~\bibnamefont
  {Koch}}, \bibinfo {author} {\bibfnamefont {A.}~\bibnamefont {Majumder}}, \
  and\ \bibinfo {author} {\bibfnamefont {J.}~\bibnamefont {Randrup}},\ }\href
  {\doibase 10.1103/PhysRevLett.95.182301} {\bibfield  {journal} {\bibinfo
  {journal} {Phys. Rev. Lett.}\ }\textbf {\bibinfo {volume} {95}},\ \bibinfo
  {pages} {182301} (\bibinfo {year} {2005})},\ \Eprint
  {http://arxiv.org/abs/nucl-th/0505052} {arXiv:nucl-th/0505052 [nucl-th]}
  \BibitemShut {NoStop}%
\bibitem [{\citenamefont {Bazavov}\ \emph {et~al.}(2013)\citenamefont
  {Bazavov}, \citenamefont {Ding}, \citenamefont {Hegde}, \citenamefont
  {Kaczmarek}, \citenamefont {Karsch} \emph {et~al.}}]{Bazavov:2013dta}%
  \BibitemOpen
  \bibfield  {author} {\bibinfo {author} {\bibfnamefont {A.}~\bibnamefont
  {Bazavov}}, \bibinfo {author} {\bibfnamefont {H.~T.}\ \bibnamefont {Ding}},
  \bibinfo {author} {\bibfnamefont {P.}~\bibnamefont {Hegde}}, \bibinfo
  {author} {\bibfnamefont {O.}~\bibnamefont {Kaczmarek}}, \bibinfo {author}
  {\bibfnamefont {F.}~\bibnamefont {Karsch}},  \emph {et~al.},\ }\href
  {\doibase 10.1103/PhysRevLett.111.082301} {\bibfield  {journal} {\bibinfo
  {journal} {Phys. Rev. Lett.}\ }\textbf {\bibinfo {volume} {111}},\ \bibinfo
  {pages} {082301} (\bibinfo {year} {2013})},\ \Eprint
  {http://arxiv.org/abs/1304.7220} {arXiv:1304.7220 [hep-lat]} \BibitemShut
  {NoStop}%
\bibitem [{\citenamefont {Kitazawa}\ and\ \citenamefont
  {Asakawa}(2012)}]{Kitazawa:2012at}%
  \BibitemOpen
  \bibfield  {author} {\bibinfo {author} {\bibfnamefont {M.}~\bibnamefont
  {Kitazawa}}\ and\ \bibinfo {author} {\bibfnamefont {M.}~\bibnamefont
  {Asakawa}},\ }\href {\doibase 10.1103/PhysRevC.86.024904} {\bibfield
  {journal} {\bibinfo  {journal} {Phys. Rev. C}\ }\textbf {\bibinfo {volume}
  {86}},\ \bibinfo {pages} {024904} (\bibinfo {year} {2012})},\ \bibinfo {note}
  {[Erratum: Phys.Rev.C 86, 069902 (2012)]},\ \Eprint
  {http://arxiv.org/abs/1205.3292} {arXiv:1205.3292 [nucl-th]} \BibitemShut
  {NoStop}%
\bibitem [{\citenamefont {Bzdak}\ and\ \citenamefont
  {Koch}(2012)}]{Bzdak:2012ab}%
  \BibitemOpen
  \bibfield  {author} {\bibinfo {author} {\bibfnamefont {A.}~\bibnamefont
  {Bzdak}}\ and\ \bibinfo {author} {\bibfnamefont {V.}~\bibnamefont {Koch}},\
  }\href {\doibase 10.1103/PhysRevC.86.044904} {\bibfield  {journal} {\bibinfo
  {journal} {Phys. Rev.}\ }\textbf {\bibinfo {volume} {C86}},\ \bibinfo {pages}
  {044904} (\bibinfo {year} {2012})},\ \Eprint {http://arxiv.org/abs/1206.4286}
  {arXiv:1206.4286 [nucl-th]} \BibitemShut {NoStop}%
\bibitem [{\citenamefont {Bzdak}\ and\ \citenamefont
  {Koch}(2015)}]{Bzdak:2013pha}%
  \BibitemOpen
  \bibfield  {author} {\bibinfo {author} {\bibfnamefont {A.}~\bibnamefont
  {Bzdak}}\ and\ \bibinfo {author} {\bibfnamefont {V.}~\bibnamefont {Koch}},\
  }\href {\doibase 10.1103/PhysRevC.91.027901} {\bibfield  {journal} {\bibinfo
  {journal} {Phys. Rev.}\ }\textbf {\bibinfo {volume} {C91}},\ \bibinfo {pages}
  {027901} (\bibinfo {year} {2015})},\ \Eprint {http://arxiv.org/abs/1312.4574}
  {arXiv:1312.4574 [nucl-th]} \BibitemShut {NoStop}%
\bibitem [{\citenamefont {Luo}(2015)}]{Luo:2014rea}%
  \BibitemOpen
  \bibfield  {author} {\bibinfo {author} {\bibfnamefont {X.}~\bibnamefont
  {Luo}},\ }\href {\doibase 10.1103/PhysRevC.91.034907} {\bibfield  {journal}
  {\bibinfo  {journal} {Phys. Rev.}\ }\textbf {\bibinfo {volume} {C91}},\
  \bibinfo {pages} {034907} (\bibinfo {year} {2015})},\ \Eprint
  {http://arxiv.org/abs/1410.3914} {arXiv:1410.3914 [physics.data-an]}
  \BibitemShut {NoStop}%
\bibitem [{\citenamefont {Bzdak}\ \emph {et~al.}(2016)\citenamefont {Bzdak},
  \citenamefont {Holzmann},\ and\ \citenamefont {Koch}}]{Bzdak:2016qdc}%
  \BibitemOpen
  \bibfield  {author} {\bibinfo {author} {\bibfnamefont {A.}~\bibnamefont
  {Bzdak}}, \bibinfo {author} {\bibfnamefont {R.}~\bibnamefont {Holzmann}}, \
  and\ \bibinfo {author} {\bibfnamefont {V.}~\bibnamefont {Koch}},\ }\href
  {\doibase 10.1103/PhysRevC.94.064907} {\bibfield  {journal} {\bibinfo
  {journal} {Phys. Rev.}\ }\textbf {\bibinfo {volume} {C94}},\ \bibinfo {pages}
  {064907} (\bibinfo {year} {2016})},\ \Eprint
  {http://arxiv.org/abs/1603.09057} {arXiv:1603.09057 [nucl-th]} \BibitemShut
  {NoStop}%
\bibitem [{\citenamefont {Adamczewski-Musch}\ \emph {et~al.}(2020)\citenamefont
  {Adamczewski-Musch} \emph {et~al.}}]{Adamczewski-Musch:2020slf}%
  \BibitemOpen
  \bibfield  {author} {\bibinfo {author} {\bibfnamefont {J.}~\bibnamefont
  {Adamczewski-Musch}} \emph {et~al.} (\bibinfo {collaboration} {HADES}),\
  }\href {\doibase 10.1103/PhysRevC.102.024914} {\bibfield  {journal} {\bibinfo
   {journal} {Phys. Rev. C}\ }\textbf {\bibinfo {volume} {102}},\ \bibinfo
  {pages} {024914} (\bibinfo {year} {2020})},\ \Eprint
  {http://arxiv.org/abs/2002.08701} {arXiv:2002.08701 [nucl-ex]} \BibitemShut
  {NoStop}%
\bibitem [{\citenamefont {Adam}\ \emph {et~al.}(2020)\citenamefont {Adam} \emph
  {et~al.}}]{Adam:2020unf}%
  \BibitemOpen
  \bibfield  {author} {\bibinfo {author} {\bibfnamefont {J.}~\bibnamefont
  {Adam}} \emph {et~al.} (\bibinfo {collaboration} {STAR}),\ }\href@noop {} {\
  (\bibinfo {year} {2020})},\ \Eprint {http://arxiv.org/abs/2001.02852}
  {arXiv:2001.02852 [nucl-ex]} \BibitemShut {NoStop}%
\bibitem [{\citenamefont {Kitazawa}(2016)}]{Kitazawa:2016awu}%
  \BibitemOpen
  \bibfield  {author} {\bibinfo {author} {\bibfnamefont {M.}~\bibnamefont
  {Kitazawa}},\ }\href {\doibase 10.1103/PhysRevC.93.044911} {\bibfield
  {journal} {\bibinfo  {journal} {Phys. Rev.}\ }\textbf {\bibinfo {volume}
  {C93}},\ \bibinfo {pages} {044911} (\bibinfo {year} {2016})},\ \Eprint
  {http://arxiv.org/abs/1602.01234} {arXiv:1602.01234 [nucl-th]} \BibitemShut
  {NoStop}%
\bibitem [{\citenamefont {Nonaka}\ \emph {et~al.}(2018)\citenamefont {Nonaka},
  \citenamefont {Kitazawa},\ and\ \citenamefont {Esumi}}]{Nonaka:2018mgw}%
  \BibitemOpen
  \bibfield  {author} {\bibinfo {author} {\bibfnamefont {T.}~\bibnamefont
  {Nonaka}}, \bibinfo {author} {\bibfnamefont {M.}~\bibnamefont {Kitazawa}}, \
  and\ \bibinfo {author} {\bibfnamefont {S.}~\bibnamefont {Esumi}},\ }\href
  {\doibase 10.1016/j.nima.2018.08.013} {\bibfield  {journal} {\bibinfo
  {journal} {Nucl. Instrum. Meth.}\ }\textbf {\bibinfo {volume} {A906}},\
  \bibinfo {pages} {10} (\bibinfo {year} {2018})},\ \Eprint
  {http://arxiv.org/abs/1805.00279} {arXiv:1805.00279 [physics.data-an]}
  \BibitemShut {NoStop}%
\bibitem [{\citenamefont {Adam}\ \emph {et~al.}(2019)\citenamefont {Adam} \emph
  {et~al.}}]{Adam:2019xmk}%
  \BibitemOpen
  \bibfield  {author} {\bibinfo {author} {\bibfnamefont {J.}~\bibnamefont
  {Adam}} \emph {et~al.} (\bibinfo {collaboration} {STAR}),\ }\href {\doibase
  10.1103/PhysRevC.100.014902} {\bibfield  {journal} {\bibinfo  {journal}
  {Phys. Rev. C}\ }\textbf {\bibinfo {volume} {100}},\ \bibinfo {pages}
  {014902} (\bibinfo {year} {2019})},\ \Eprint
  {http://arxiv.org/abs/1903.05370} {arXiv:1903.05370 [nucl-ex]} \BibitemShut
  {NoStop}%
\bibitem [{eff()}]{effcorr:github}%
  \BibitemOpen
  \href@noop {} {}\bibinfo {note}
  {\href{https://github.com/vlvovch/off-diagonal-efficiency}{https://github.com/vlvovch/off-diagonal-efficiency}
  [Online; accessed 08-January-2021]}\BibitemShut {NoStop}%
\bibitem [{\citenamefont {Nonaka}\ \emph {et~al.}(2017)\citenamefont {Nonaka},
  \citenamefont {Kitazawa},\ and\ \citenamefont {Esumi}}]{Nonaka:2017kko}%
  \BibitemOpen
  \bibfield  {author} {\bibinfo {author} {\bibfnamefont {T.}~\bibnamefont
  {Nonaka}}, \bibinfo {author} {\bibfnamefont {M.}~\bibnamefont {Kitazawa}}, \
  and\ \bibinfo {author} {\bibfnamefont {S.}~\bibnamefont {Esumi}},\ }\href
  {\doibase 10.1103/PhysRevC.95.064912} {\bibfield  {journal} {\bibinfo
  {journal} {Phys. Rev.}\ }\textbf {\bibinfo {volume} {C95}},\ \bibinfo {pages}
  {064912} (\bibinfo {year} {2017})},\ \Eprint
  {http://arxiv.org/abs/1702.07106} {arXiv:1702.07106} \BibitemShut {NoStop}%
\bibitem [{\citenamefont {Abelev}\ \emph {et~al.}(2013)\citenamefont {Abelev}
  \emph {et~al.}}]{Abelev:2012pv}%
  \BibitemOpen
  \bibfield  {author} {\bibinfo {author} {\bibfnamefont {B.}~\bibnamefont
  {Abelev}} \emph {et~al.} (\bibinfo {collaboration} {ALICE}),\ }\href
  {\doibase 10.1103/PhysRevLett.110.152301} {\bibfield  {journal} {\bibinfo
  {journal} {Phys. Rev. Lett.}\ }\textbf {\bibinfo {volume} {110}},\ \bibinfo
  {pages} {152301} (\bibinfo {year} {2013})},\ \Eprint
  {http://arxiv.org/abs/1207.6068} {arXiv:1207.6068 [nucl-ex]} \BibitemShut
  {NoStop}%
\end{thebibliography}%

\end{document}